\documentclass[amsmath,amssymb,aps,prl,twocolumn,nofootinbib,preprintnumbers]{revtex4}
\newcommand{\be}{\begin{equation}}
\newcommand{\bea}{\begin{eqnarray}}
\newcommand{\eea}{\end{eqnarray}}
\newcommand{\ba}{\begin{array}}
\newcommand{\ea}{\end{array}}
\newcommand{\ee}{\end{equation}}

\begin{document}
\title{Replicated Entanglement Entropy }
\author{Amir Esmaeil Mosaffa$\footnote{mosaffa@theory.ipm.ac.ir}$}
\affiliation{\,\, {\sl School of Particles and Accelerators,\\ Institute for Research in Fundamental Sciences (IPM), \\
		P.O. Box 19395-5531, Tehran, Iran}}

\begin{abstract} 
We introduce ``Replicated Entanglement Entropy (REE)" as the  entanglement entropy of a subspace in a replicated theory. We calculate this quantity by replicating the original theory in two steps along the same entangling region and taking the proper limit on the second replica number. The quantity has a clear physical meaning, a thermodynamic interpretation and a holographic dual.

\end{abstract}
\preprint{}

\maketitle

\underline{\textit{Introduction}}\\

In various physical situations one is often interested in measurements carried out by an observer who has a partial access to the total degrees of freedom. Quantum mechanically this becomes even more interesting as such experiments reveal the quantum structure of states. The relevant operator corresponding to such measurements is the \textit{reduced density} operator denoted by $\rho$. This is obtained by starting with the density operator of the whole system and taking a partial trace over the unaccessible degrees of freedom.

In an ideal case one knows all the eigenvalues of $\rho$, known as the \textit{entanglement spectrum}. Equivalently, one may look for a sequence of quantities known as $R\acute{e}ney\ entropies$ defined by
\be
S_n=\frac{1}{1-n}\log \rm{tr}\rho^n\ ,
\ee
where $n$ is a positive integer known as the replica number and ``tr" stands for trace. The first entry in this sequence, $n\rightarrow1$, is called the \textit{entanglement entropy} of the abovementioned subsystem and is denoted by $S$. 

Here we are interested in a continuum field theory in $d+1$ spacetime dimensions and our subsystem consists of a $d$ dimensional spatial subspace denoted by $\cal{A}$ and called entangling region. Replication is produced by providing $n$ copies of the spacetime manifold, cutting each along $\cal{A}$ 
and gluing them in a cyclic order along the cuts (see\cite{Calabrese:2004eu,Calabrese:2005zw,Calabrese:2009qy}). We will denote by $Z$ the partition function of the original theory and by $Z^{\cal{A}}_n$ that of the replicated one. In terms of these quantities, the $n$-th
$\rm{R\acute{e}ney\ entropy}$ is written as
\be
\label{na}
S^{\cal{A}}_n=\frac{1}{1-n}\left[ \log Z^{\cal{A}}_n-n\log Z\right]. 
\ee

In the context of gauge/gravity duality \cite{Maldacena:1997re, Gubser:1998bc, Witten:1998qj}, entanglement entropy has a distinct role as it has a holographic description. For theories which have a gravitational dual, Ryu and Takayanagi (RT) have proposed that $S^{\cal{A}}$ of the subspace $\cal{A}$ is
proportional to the area of a codimention 2 minimal surface in the gravitational background \cite{Ryu:2006bv, Ryu:2006ef}. It is holologeous to $\cal{A}$ on the asymptotic boundary where the field theory lives.

There has also been progress in finding holographic duals for $\rm{R\acute{e}ney\ entropies}$ \cite{Headrick:2010zt, Hung:2011nu, Fursaev:2012mp, Faulkner:2013yia, Galante:2013wta, Belin:2013dva, Barrella:2013wja, Chen:2013dxa}. In a recent work \cite{Dong:2016fnf}, a modification of $\rm{R\acute{e}ney\ entropy}$ has been introduced and shown to have a holographic description in terms of an area law, very much in the spirit of the RT proposal. In this work we introduce the same quanity of \cite{Dong:2016fnf} but as the entanglement entropy in a replicated theory, which we call Replicated Entanglement Entropy (REE). Being an entanglement, rather than a $\rm{R\acute{e}ney}$ entropy, the RT proposal gives us the desired holographic description.

We start with the $n$-replicated theory and wish to calculate the entanglement entropy of a certain subspace $\cal{B}$. This amounts to replicating once more by $m$ and taking the limit $m\rightarrow1$. In case of an existing gravitational dual for the theory, the quantity we calculate can be expressed as the area of its corresponding RT curve. This time, though, the gravitational background has the $n$-replicated field theory
on its boundary.

Furthermore, in the special case of $\cal{B}$=$\cal{A}$, the resulting entropy can be written in terms of the $n$-th $\rm{R\acute{e}ney\ entropy}$ and its derivative and will turn out to coincide with the modified $\rm{R\acute{e}ney\ entropy}$ of \cite{Dong:2016fnf}. \\

\underline{\textit{Replicated Entanglement Entropy}}\\

Suppose we have an $n$-replica of a theory along an entangling region $\cal{A}$. We now choose a second entangling region $\cal{B}$
and replicate once more by $m$ times. The resulting $\rm{R\acute{e}ney\ entropy}$ which we denote by $(S^{\cal{A}}_n)^{\cal{B}}_m$ is
\be
(S^{\cal{A}}_n)^{\cal{B}}_m=\frac{1}{1-m}\left[ \log (Z^{\cal{A}}_n)^{\cal{B}}_m-m\log Z^{\cal{A}}_n\right]\ ,
\ee
where $(Z^{\cal{A}}_n)^{\cal{B}}_m$ represents the partition function after double replication. One can now try to take the limit of $m\rightarrow1$ and find the entanglement entropy of $\cal{B}$ on the replicated theory, $(S^{\cal{A}}_n)^{\cal{B}}$.
Whatever the result is, as long as we have a gravitational description for the replicated theory and according to the Ryu-Takayanagi prescription, this quantity will be described by a minimal surface in the bulk space. This space has the replicated theory living on its asymptotics and the minimal surface ends on the entangling region $\cal{B}$ on the boundary.

Now suppose that we cut the $n$-replica along one of its $n$ copies of $\cal{A}$ and replicate it $m$ times, that is, choose $\cal{B}$ to be the original entangling region $\cal{A}$. It is not hard to convince oneself that the resulting theory will be the $nm$-replica of the original one along $\cal{A}$ and that 
\be
(Z^{\cal{A}}_n)^{\cal{A}}_m=Z^{\cal{A}}_{nm}\ .
\ee
However, $(S^{\cal{A}}_n)^{\cal{A}}_m\ne S^{\cal{A}}_{nm}$. That is, the $m$-th $\rm{R\acute{e}ney\ entropy}$ of the $n$-replicated theory is not the same as the $nm$-th $\rm{R\acute{e}ney\ entropy}$ of the original theory. Rather, it is given by

\be
\label{nama}
(S^{\cal{A}}_n)^{\cal{A}}_m=\frac{1}{1-m}\left[ \log Z^{\cal{A}}_{nm}-m\log Z^{\cal{A}}_n\right]\ ,
\ee
whereas
\be
\label{nma}
S^{\cal{A}}_{nm}=\frac{1}{1-nm}\left[ \log Z^{\cal{A}}_{nm}-nm\log Z\right]. 
\ee
We can use equations (\ref{na}) and (\ref{nma}) to write (\ref{nama}) as
\be
(S^{\cal{A}}_n)^{\cal{A}}_m=\frac{1}{1-m}\left[ (1-nm)S^{\cal{A}}_{nm}-m(1-n)S^{\cal{A}}_n\right] .
\ee

To find the entanglement entropy we now take the 
$m\rightarrow1$ limit of the above equation by setting $m=1+\epsilon$ and taking the limit $\epsilon\rightarrow0$.
We find (we drop the index $\cal{A}$)
\be
(S_n)_{1+\epsilon}\approx\frac{1}{\epsilon}\left[\epsilon S_n+n(n-1)\delta S_n\right]\ ,
\ee
and thus
\be
\label{tilde}
\tilde{S}_n\equiv\lim_{m\rightarrow1}(S_n)_m=n^2\partial_n\left( \frac{n-1}{n}S_n\right). 
\ee
This is the quantity that was recently proposed in \cite{Dong:2016fnf} as a modification to $\rm{R\acute{e}ney\ entropy}$. We see here that it is in fact an entanglement entropy but in a replicated theory.
The construction makes it obvious that the $n\rightarrow1$ limit reproduces the usual entanglement entropy. Furthemore it was also shown in \cite{Dong:2016fnf} that $\tilde{S}_n$ has a dual description by holography. Here, this fact is a natural consequence of the renowned Ryu-Takayanagi proposal for the holographic dual of entanglement entropy. \\

\underline{\textit{Holographic dual of $\tilde{S}_n$}} \\

According to RT prescription, to find the holographic dual of
$\tilde{S}_n$ we must calculate the area of a codimension 2 minimal surface in a certain $d+2$ dimensional gravitational background. In this case such a background must have an asymptotic boundary on which the replicated theory resides. We follow \cite{Dong:2016fnf} to denote such a gravitational solution by $B_n$. 

The $n$-replicated manifold on the asymptotic boundary has a conical singularity along $\partial\cal{A}$, the boundary of the entangling region. This cone has a deficit angle $2\pi(n-1)$. It turns out that the dual gravitational solution will be smooth in the bulk \cite{Headrick:2010zt}. 

Alternatively, one can make use of the $\mathbb{Z}_n$ replica symmetry to replace the replicated boundary manifold with a single copy of the original one. Instead, one should now make an orbifold of the bulk solution $\hat{B}_n=B_n/\mathbb{Z}_n$ to produce the dual gravitational side. The $\mathbb{Z}_n$ action has fixed points in the bulk whose locus constitute a codimension 2 surface. As a result the bulk solution will have a conical singularity along this surface and the corresponding deficit angle will be $2\pi(n-1)/n$ \cite{Lewkowycz:2013nqa} (see also \cite{Haehl:2014zoa}).

In \cite{Dong:2016fnf} the latter approach has been used to find the holographic dual of $\tilde{S}_n$
\be
\label{dong}
\tilde{S}_n=\dfrac{\rm{Area (Cosmic\ Brane_\mathit{n})}}{4G_N}\ ,
\ee
where the cosmic brane's dynamic is governed by the Nambu-Gotto action. It backreacts on the geometry and produces the mentioned conical singularity. Its profile is found by solving for the saddle point of the total action of bulk and brane. 

Here, our construction of $\tilde{S}_n$ as an entanglement entropy suggests the former approach, i.e. the one with a replicated boundary and a smooth bulk, as a natural holographic setup. By the standard RT prescription, this amounts to finding the solution $B_n$ and calculating the area of the codimension 2 minimal surface homologeous to $\cal{A}$\footnote{We choose one of the $n$ copies of $\cal{A}$ on the boundary.}
\be
\label{mine}
\tilde{S}_n=\dfrac{\rm{Area (Minimal\ Surface)}}{4G_N}\ .
\ee

To sum up, the proposal in (\ref{dong}) works with a smooth boundary and a singular bulk $\hat{B}_n$ whereas in (\ref{mine})
we work with a singular boundary and a smooth bulk $B_n$.

We give an example in the following.\\

\underline{\textit{An Example}}\\

We test (\ref{mine}) in $AdS_3$, parametrized by $(z,x,\tau)$. The boundary is a replicated two dimensional CFT. We focus on one of the $n$ replicas with coordinates $(x,\tau)$ and  choose the entangling region to be a spatial interval with ends at $(-a,0)$ and $(a,0)$. The smooth bulk solution $B_n$ in this case can be found by methods developed in \cite{Skenderis:1999nb} (see appendix C of \cite{Hung:2011nu} for details). The bulk metric at the $t=0$ slice is found as
\be
ds^2|_{t=0}=\frac{dz^2}{z^2}+\left( \frac{1}{z^2}-\frac{a^2(1-1/n^2)}{(x^2-a^2)^2}\right) dx^2\ .
\ee
This is the usual $AdS$ metric with an additional term\footnote{We take the $AdS$ radius to be 1}.
To find the minimal surface one should solve the geodesic equation with the above metric. As an approximation we consider the new term to be small. This can be achieved by assuming that $n\approx1$. One may also note that this term is only important when $x\approx\pm a$, that is, when the geodesic curve touches the boundary. We thus keep $n$ arbitrary and assume that the additional term does not change the shape of the geodesic curve and only changes its length.

The geodesic is given by a semicircle $x^2+z^2=a^2$ and its length is given by
\be
L=2\int_{\delta}^{a}\frac{dz}{z}\sqrt{1+\frac{z^2}{a^2-z^2}\left( 1-\frac{z^2a^2(1-1/n^2)}{(x^2-a^2)^2}\right)}
\ee	
where $\delta$ is the UV cutoff. Upon integration we find
\be
L=\frac{2}{n} \log{\frac{2a}{\delta}}\ .
\ee
Once we substitute this in (\ref{mine})  and recalling that $\frac{3}{2G_N}=c$, the central charge of field theory, we find
\be
\label{holres}
\tilde{S}_n=\dfrac{\rm{Area (Minimal\ Surface)}}{4G_N}=\frac{1}{n}\ \frac{c}{3}\log{\frac{2a}{\delta}}\ .
\ee
On the other hand we know the $\rm{R\acute{e}ney\ entropy}$ of an interval to be
\be
S_n=\frac{c}{6}(1+\frac{1}{n})\log{\frac{2a}{\delta}}\ .
\ee
Plugging this in the definition of $\tilde{S}_n$ we find
\be
\tilde{S}_n=n^2\partial_n\left( \frac{n-1}{n}S_n\right)=\frac{1}{n}\ \frac{c}{3}\log{\frac{2a}{\delta}}\ .
\ee
Therefore the holographic result as obtained in (\ref{holres}) agrees with the direct computation in field theory. Curious enough
despite an approximate calculation in the bulk we get the exact answer. More examples should be examined for a better insight.\\

\underline{\textit{Thermodynamics of $\tilde{S}_n$}}\\

A very insightful description for entanglement entropy of spherical regions was developed in \cite{Casini:2011kv} in terms of thermal entropy. The analysis allows to write the entanglement entropy for such regions in terms of the thermodynamic free energy $F(T)$
\be
S=-\frac{\partial F}{\partial T}\ ,
\ee
where the temperature is inversely proportional to the radius of the entangling sphere. In \cite{Hung:2011nu} further thermodynamic connections were found for the  $\rm{R\acute{e}ney\ entropy}$ of spherical regions 
\be
\label{renfree}
S_n=\frac{n}{1-n}[F(1)-F(1/n)]\ ,
\ee
where we have taken the  temperature to be 1. 
The free energy $F(1/n)$ is associated to the replicated theory as shown in \cite{Hung:2011nu}. We see that the thermal entropy resulting from this free energy is the replicated entanglement entropy
\be
\tilde{S}_n=n^2\partial_n\left( \frac{n-1}{n}S_n\right)=-\frac{\partial F(1/n)}{\partial (1/n)}\ .
\ee
The conclusion is that the replicated entanglement entropy for the sphere can also be understood in thermal terms with a temperature $1/n$. Similar arguments were given in \cite{Dong:2016fnf}.\\

\underline{\textit{Final Remarks}}\\

Replicated Entanglement Entropy may prove to be useful in finding a geometric holographic description for $\rm{R\acute{e}ney\ entropy}$.
It will be interesting to perform the analysis of \cite{Casini:2011kv} for a direct thermal description in the replicated theory. Also, the two holographic descriptions in (\ref{dong}) and (\ref{mine}) require the cosmic brane in $\hat{B_n}$ and the minimal surface in $B_n$ to have the same area. This may have interesting consequences.\\

I would like to thank Hossein Seyedi for discussions.


\begin{thebibliography}{10}
  
\bibitem{Calabrese:2004eu} 
  P.~Calabrese and J.~L.~Cardy,
  ``Entanglement entropy and quantum field theory,''
  J.\ Stat.\ Mech.\  {\bf 0406}, P06002 (2004)
  [hep-th/0405152].
  
\bibitem{Calabrese:2005zw} 
  P.~Calabrese and J.~L.~Cardy,
  ``Entanglement entropy and quantum field theory: A Non-technical introduction,''
  Int.\ J.\ Quant.\ Inf.\  {\bf 4}, 429 (2006)
  [quant-ph/0505193].
  
\bibitem{Calabrese:2009qy} 
  P.~Calabrese and J.~Cardy,
  ``Entanglement entropy and conformal field theory,''
  J.\ Phys.\ A A {\bf 42}, 504005 (2009)
  [arXiv:0905.4013 [cond-mat.stat-mech]].
  
  \bibitem{Maldacena:1997re} 
  J.~M.~Maldacena,
  ``The Large N limit of superconformal field theories and supergravity,''
  Int.\ J.\ Theor.\ Phys.\  {\bf 38}, 1113 (1999)
  [Adv.\ Theor.\ Math.\ Phys.\  {\bf 2}, 231 (1998)]
  doi:10.1023/A:1026654312961
  [hep-th/9711200].
  
  \bibitem{Gubser:1998bc} 
  S.~S.~Gubser, I.~R.~Klebanov and A.~M.~Polyakov,
  ``Gauge theory correlators from noncritical string theory,''
  Phys.\ Lett.\ B {\bf 428}, 105 (1998)
  doi:10.1016/S0370-2693(98)00377-3
  [hep-th/9802109].
  
  \bibitem{Witten:1998qj} 
  E.~Witten,
  ``Anti-de Sitter space and holography,''
  Adv.\ Theor.\ Math.\ Phys.\  {\bf 2}, 253 (1998)
  [hep-th/9802150].
  
  
  
\bibitem{Ryu:2006bv} 
  S.~Ryu and T.~Takayanagi,
  ``Holographic derivation of entanglement entropy from AdS/CFT,''
  Phys.\ Rev.\ Lett.\  {\bf 96}, 181602 (2006)
  [hep-th/0603001].
  
    
\bibitem{Ryu:2006ef} 
  S.~Ryu and T.~Takayanagi,
  ``Aspects of Holographic Entanglement Entropy,''
  JHEP {\bf 0608}, 045 (2006)
  [hep-th/0605073].
  
 \bibitem{Headrick:2010zt} 
 M.~Headrick,
 ``Entanglement Renyi entropies in holographic theories,''
 Phys.\ Rev.\ D {\bf 82}, 126010 (2010)
 doi:10.1103/PhysRevD.82.126010
 [arXiv:1006.0047 [hep-th]].
  
  \bibitem{Hung:2011nu} 
  L.~Y.~Hung, R.~C.~Myers, M.~Smolkin and A.~Yale,
  ``Holographic Calculations of Renyi Entropy,''
  JHEP {\bf 1112}, 047 (2011)
  doi:10.1007/JHEP12(2011)047
  [arXiv:1110.1084 [hep-th]].
  
  \bibitem{Fursaev:2012mp} 
  D.~V.~Fursaev,
  ``Entanglement Renyi Entropies in Conformal Field Theories and Holography,''
  JHEP {\bf 1205}, 080 (2012)
  doi:10.1007/JHEP05(2012)080
  [arXiv:1201.1702 [hep-th]].
  
  \bibitem{Faulkner:2013yia} 
  T.~Faulkner,
  ``The Entanglement Renyi Entropies of Disjoint Intervals in AdS/CFT,''
  arXiv:1303.7221 [hep-th].
  
  \bibitem{Galante:2013wta} 
  D.~A.~Galante and R.~C.~Myers,
  ``Holographic Renyi entropies at finite coupling,''
  JHEP {\bf 1308}, 063 (2013)
  doi:10.1007/JHEP08(2013)063
  [arXiv:1305.7191 [hep-th]].
  
  \bibitem{Belin:2013dva} 
  A.~Belin, A.~Maloney and S.~Matsuura,
  JHEP {\bf 1312}, 050 (2013)
  doi:10.1007/JHEP12(2013)050
  [arXiv:1306.2640 [hep-th]].
  
  \bibitem{Barrella:2013wja} 
  T.~Barrella, X.~Dong, S.~A.~Hartnoll and V.~L.~Martin,
  ``Holographic entanglement beyond classical gravity,''
  JHEP {\bf 1309}, 109 (2013)
  doi:10.1007/JHEP09(2013)109
  [arXiv:1306.4682 [hep-th]].
  
  \bibitem{Chen:2013dxa} 
  B.~Chen, J.~Long and J.~j.~Zhang,
  JHEP {\bf 1404}, 041 (2014)
  doi:10.1007/JHEP04(2014)041
  [arXiv:1312.5510 [hep-th]].
  
  \bibitem{Dong:2016fnf} 
  X.~Dong,
  ``An Area-Law Prescription for Holographic Renyi Entropies,''
  arXiv:1601.06788 [hep-th].
 
 \bibitem{Lewkowycz:2013nqa} 
 A.~Lewkowycz and J.~Maldacena,
 ``Generalized gravitational entropy,''
 JHEP {\bf 1308}, 090 (2013)
 doi:10.1007/JHEP08(2013)090
 [arXiv:1304.4926 [hep-th]].
 
 \bibitem{Haehl:2014zoa} 
 F.~M.~Haehl, T.~Hartman, D.~Marolf, H.~Maxfield and M.~Rangamani,
 ``Topological aspects of generalized gravitational entropy,''
 JHEP {\bf 1505}, 023 (2015)
 doi:10.1007/JHEP05(2015)023
 [arXiv:1412.7561 [hep-th]].
 
 \bibitem{Skenderis:1999nb} 
 K.~Skenderis and S.~N.~Solodukhin,
 Phys.\ Lett.\ B {\bf 472}, 316 (2000)
 doi:10.1016/S0370-2693(99)01467-7
 [hep-th/9910023].
 
 \bibitem{Casini:2011kv} 
 H.~Casini, M.~Huerta and R.~C.~Myers,
 JHEP {\bf 1105}, 036 (2011)
 doi:10.1007/JHEP05(2011)036
 [arXiv:1102.0440 [hep-th]].
 
 
 
  
 
  
\end{thebibliography}
\end{document}